\documentclass{revtex4}

\usepackage{graphicx}
\begin{document}
% You should use BibTeX and revtex.bst for references
% \bibliographystyle{revtex}

\preprint{SLAC-PUB-9033}

\title{The LCDROOT analysis package\footnote{Work supported by
Department of Energy contract  DE--AC03--76SF00515.}}

\author{Toshinori Abe}
\email[]{toshi@SLAC.Stanford.EDU}
\affiliation{
Stanford Linear Accelerator Center, Stanford CA 94309, USA
}

\author{Masako Iwasaki}
\email[]{masako@SLAC.Stanford.EDU}
\affiliation{
Department of Physics, University of Oregon, Eugene,
OR 97403-1274, USA
}

\date{\today}

\begin{abstract}
The North American Linear Collider Detector group 
has developed simulation and analysis program packages.
LCDROOT is one of the packages, 
and is based on ROOT and the C++ programing language
to maximally benefit from object oriented programming techniques.
LCDROOT is constantly improved and 
now has a new topological vertex finder, ZVTOP3.
In this proceeding, the features of the LCDROOT simulation are briefly
described.
\end{abstract}

\maketitle

%
% Introduction
%
\section{Introduction}
For various studies for future Linear Collider 
experiments,
the North American Linear Collider Detector group (LCD)
has developed simulation and analysis program packages.
LCDROOT is one of the packages
for event generation, fast and full detector simulations, 
and event analyses.
In LCDROOT,
the simulators are designed to have the flexibility of changing any component
of the detector parameters, i.e, geometry and performance,
to study detector requirements in detail.

LCDROOT is based on ROOT~\cite{Brun:1997pa} 
which is an object oriented framework,
and the C++ programming language.
In the ROOT framework the basic utilities and services, 
such as I/O and 3D graphics, 
are provided.
ROOT provides a large selection of HEP-specific utilities
such as histogramming and fitting.
ROOT also has a C++ command interpreter.
Due to these very powerful features,
a lot of experimented groups use ROOT.
Currently we support LCDROOT
on AIX, Linux, Sun and Windows platforms.

%
% Generator
%
\section{Event generator}
For event generation, we support PANDORA~\cite{Peskin:1999hh}
with an interface using
PYTHIA~\cite{Sjostrand:1994yb,Sjostrand:2000wi}
and TAUOLA~\cite{Jadach:1993hs}, for parton showering, hadronization,
and tau decays.
The generator includes beam polarization, beamstrahlung, 
and initial state radiation,
which are very important features for
future Linear Collider studies.
PANDORA is written in C++ and can be directly
handled within LCDROOT.
Hence PANDORA is used as our main generator.
We also support the PYTHIA~\cite{Sjostrand:1994yb,Sjostrand:2000wi}
event generator with a C++ interface class.
In order to support other generator outputs, 
LCDROOT also handles HEPEVT common blocks
using the FNAL StdHep I/O package~\cite{STDHEP}.

%
% Simulator
%
\section{Simulation and Event Reconstruction}
LCDROOT supports both fast and full detector simulations.
The fast simulator is designed to provide a fast 
and flexible physics analysis environment, 
while the full simulator is for 
detailed detector studies.
Detector geometry is specified by a text file translated from XML,
and geometry parameters are easily changed in LCDROOT.

The fast simulation is based on parametrized position and energy smearing, 
and it makes tracks and clusters directly from the generated particle
information.
In the fast simulation, 
charged particles within the magnetic field follow helical trajectories, 
and their momenta and positions are smeared 
according to their error matrix.
%Here the charged tracks are expressed by 5 parameters.
The error matrices are given by a look-up table method based on momentum
and $\cos\theta$ of charged particles.
The error matrices include off-diagonal elements to give added realism.

Electrons, photons, and hadrons produce clusters in the electromagnetic 
(EM) and hadronic (HAD) calorimeters.
Here, one cluster is made from one particle,
and energies and positions of clusters are smeared.
% We assume transverse position resolutions 
% of 1cm/$\sqrt{E}$ (electrons and photons) or 5cm/$\sqrt{E}$ (hadrons). 
To consider the detector granularity and cluster width, which is
typically a few units of Moliere radius, 
we merge the clusters when the angular separation between
clusters is less than $\theta_{max}$, where $\theta_{max}$ 
is a size of the detector granularity.

The position of the interaction position is also smeared.
We assume $\sigma_{x}$=$\sigma_{y}$=2 $\mu$m and 
$\sigma_{z}$=6 $\mu$m~\cite{Abe:2001pe}.

We use GISMO~\cite{Burnett:1993xa} for the full simulation.
The full simulation outputs simulation data in binary format
using SIO~\cite{SIO}.
LCDROOT reads the SIO binary format for event reconstruction.
Using the digitized outputs from the full simulation, 
we reconstruct charged tracks and clusters.
In the reconstruction of the full simulated data, 
we postpone the track reconstruction. Instead, we make charged tracks
by smearing, using exactly the same procedure as in the fast simulation, 
but we also apply a minimum tracker-hit cut. 
Calorimeter clusters are made by gathering the hits which are from the 
same particle. Energy and position of the cluster is obtained from 
the energy sum, and the energy-weighted average of associated
Calorimeter hits, respectively.

Comparing the full and fast simulations, the most 
significant difference is in the Calorimetry.
It is important to improve the parameterization of 
the fast simulator Calorimetry to have a
more realistic detector response, which we hope to implement 
in the near future. 

\section{Event Analysis Tools}
%
% Event analysis tools
%
For the physics analysis, there are several useful tools
in LCDROOT. 
We provide Thrust finding and 3 kinds of Jet finding
(based on JADE, JADE-E and DURHAM algorithms) programs. 
There is also an event display for LCDROOT.

In the energy flow analysis, neutral clusters are selected by 
the absence of a track and cluster association. 
For this analysis, we provide methods which extrapolate 
the particles to the cluster cylindrical radius. 

For heavy flavor tagging, we provide two kinds of topological
vertexing algorithms which are described in the next section.

\section{Heavy flavor tagging}
\begin{figure} [h]% fig 1
\includegraphics{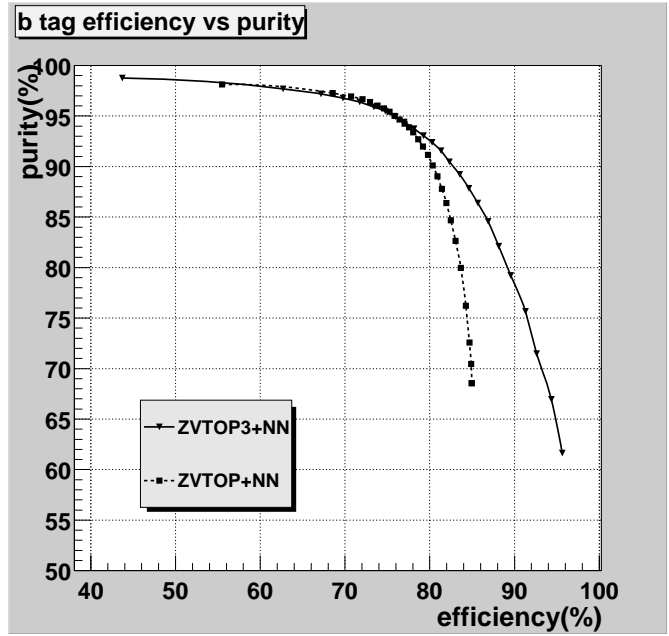}
\caption{
Performance of $b$-jet flavor tagging in $Z^0\to q\bar{q}$ events
generated at $\sqrt{s}=91.26$GeV.
}
\label{Fig:Fig1}
\end{figure}

\begin{figure} [h]% fig 2
\includegraphics{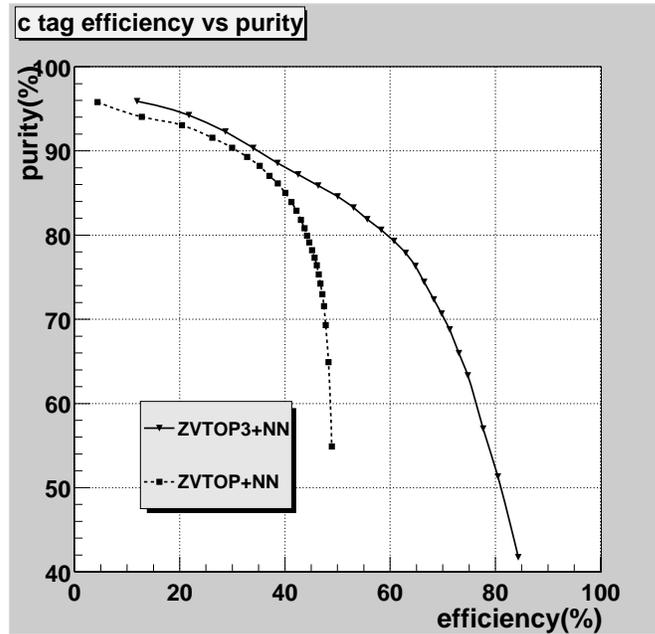}
\caption{
Performance of $c$-jet flavor tagging in $Z^0\to q\bar{q}$ events
generated at $\sqrt{s}=91.26$GeV.
}
\label{Fig:Fig2}
\end{figure}

%
% Heavy flavor tagging
%
Building on the
success of the CCD-based vertex detector (VTX) at the SLC/SLD 
experiment~\cite{Abe:1997bu,Abe:1999ky}, we strongly believe
that a CCD-based VTX will provide the optimal performance
in future Linear Collider experiments.
% Therefore, in the LCD detector design,
% we choose the CCD-based VTX.
Taking advantage of the precise 3-D spatial points
provided by such a detector, 
powerful topological vertexing techniques were developed by 
SLD~\cite{Jackson:1997sy}.
The algorithms naturally associates tracks
with the vertices where they originated and
can reconstruct a full $b$/$c$-meson decay
chain, i.e, primary, secondary, and tertiary vertices.
Using the reconstructed secondary/tertiary vertex, 
the invariant mass of the tracks associated with a decay
is used to identify jet flavor (mass tag technique~\cite{Abe:1998sb}).
This combination of the techniques gives the best heavy-flavor-jet tagging
performance in $e^+e^-$ colliding experiments at present.
%Here it should be noted that the secondary/tertiary vertex reconstruction 
%enables vertex charge information to be determined 
%which gives quark/anti-quark jet identification even 
%for neutral $B$'s.
%The secondary/tertiary vertices are reconstructed with charged tracks, 
%by searching the space points where track density functions overlap 
%in the 3D space.
We have previously ported the original topological vertexing source code, 
ZVTOP, written in Prepmort, to C++ to fit natively into the LCDROOT
simulation environment.
The main feature of ZVTOP3 is the ability to reconstruct one-prong
decay vertices, giving it better reconstruction efficiency.
During Snowmass 2001 we succeeded in implementing this functionality
into LCDROOT.
Figs.~\ref{Fig:Fig1} and \ref{Fig:Fig2} show 
$b$/$c$-quark jet tagging efficiencies
and purities comparing the performance of ZVTOP and ZVTOP3, 
clearly demonstrating the enhanced capabilities of this algorithm.
We plan to continue our efforts to improve the flavor-tagging efficiency 
and purity since it is such an essential part of any Linear Collider
physics program.


%
% Summary
%
\section{Summary}
In this report, we briefly introduce the simulation and analysis tools
based on ROOT for the LCD group. 
The tools are constantly being improved and can be
obtained via the URL,
\url{
   http://www-sldnt.slac.stanford.edu/nld/New/Docs/LCD_Root/root.htm.
}
Feedback from the users is highly welcomed.

%\begin{acknowledgments}
%\end{acknowledgments}

\bibliography{E3045}

\end{document}